\documentclass[aps,prd,showpacs,showkeys,
preprintnumbers,unsortedaddress,superscriptaddress,
twocolumn,nofootinbib]{revtex4-1}

\usepackage{amssymb}
\usepackage{amsmath}
\usepackage{bm} 
\usepackage{graphicx}
\usepackage{color}
\usepackage{xcolor}
\usepackage{dsfont}
\usepackage{enumerate}
\usepackage{hyperref}

\begin{document}

\title{An eigenvector based approach to neutrino mixing}

\date{\today}

\author{D. Aristizabal Sierra}
\email{daristizabal@ulg.ac.be}
\affiliation{\small IFPA, Dep. AGO, Universite de
      Liege, Bat B5,\\ \small \sl Sart Tilman B-4000 Liege 1,
      Belgium}

\author{I. de Medeiros Varzielas}
\email{ivo.de@udo.edu}
\affiliation{\small Facult\"at f\"ur Physik, Technische 
      Universit\"at Dortmund\\ D-44221 Dortmund, Germany}

\author{E. Houet}
\email{eric.houet@student.ulg.ac.be}
\affiliation{\small IFPA, Dep. AGO, Universite de
      Liege, Bat B5,\\ \small \sl Sart Tilman B-4000 Liege 1,
      Belgium}




\preprint{DO-TH 13/06}

\begin{abstract}
  We propose a model-independent analysis of the neutrino mass matrix
  through an expansion in terms of the eigenvectors defining the
  lepton mixing matrix, which we show can be parametrized as small
  perturbations of the tribimaximal mixing eigenvectors. This approach
  proves to be powerful and convenient for some aspects of lepton
  mixing, in particular when studying the sensitivity of the mass
  matrix elements to departures from their tribimaximal form. In terms
  of the eigenvector decomposition, the neutrino mass matrix can be
  understood as originating from a tribimaximal dominant structure
  with small departures determined by data. By implementing this
  approach to cases when the neutrino masses originate from different
  mechanisms, we show that the experimentally observed structure
  arises very naturally.  We thus claim that the observed deviations
  from the tribimaximal mixing pattern might be interpreted as a
  possible hint of a ``hybrid'' nature of the neutrino mass matrix.
 
\end{abstract}

\maketitle



There are two main approaches to describe lepton flavor mixing. One is
based on assuming the mixing is governed by a
fundamental organizing principle, such as a flavor symmetry, which 
dictates the structure of the lepton mixing
pattern and might eventually account for quark mixing as well (see
e.g. \cite{Altarelli:2010gt,King:2013eh}). The other, usually referred to as the anarchy approach,
postulates that lepton mixing originates from a random distribution of
unitary $3\times 3$ matrices \cite{Hall:1999sn}.
In either case these approaches are far from providing an ultimate
solution to the lepton flavor puzzle.
Before the striking measurements of $\theta_{13}$
\cite{An:2012eh,Ahn:2012nd}, even though global fits 
had hinted to a non-vanishing $\theta_{13}$
\cite{Fogli:2008jx}, lepton mixing was well described by the Tribimaximal
mixing (TBM) pattern \cite{Harrison:2002er} defined by
$\sin^2\theta_{12}=1/3$, $\sin^2\theta_{23}=1/2$ and
$\sin^2\theta_{13}=0$. The TBM pattern was for almost a decade a
paradigm since its regularity is very suggestive of an underlying
principle at work. With a vanishing $\theta_{13}$ now excluded at more
than $10\sigma$ \cite{Tortola:2012te} the situation has changed somewhat.
The advent of experimental data proving non-vanishing $\theta_{13}$ and
deviations of the best-fit-point values (BFPVs) of the other angles (particularly $\theta_{23}$) from their TBM values
\cite{Tortola:2012te,GonzalezGarcia:2012sz,Fogli:2012ua} has greatly motivated the search for possible
mechanisms yielding the required deviations from the TBM pattern,
which almost without exception are induced by effective operators.
In this way, flavor models unable to produce ``large'' deviations on $\theta_{13}$
from its TBM value have been ruled out, and often deviations on the TBM pattern must be sourced from
next-to-leading order non-renormalizable operators, constraining model building.

Majorana neutrino masses can be
incorporated in the standard model Lagrangian through the dimension
five effective operator ${\cal O}_5\sim LLHH$ \cite{Weinberg:1979sa},
the type-I seesaw \cite{seesaw} being the most popular and simplest
realization of this operator. Other
realizations 
have been considered as pathways
to neutrino masses but often the resulting neutrino mass
matrix is solely sourced by a single set of lepton number violating
parameters, e.g. in type-I seesaw the right-handed neutrino
masses. However, given the multiple realizations of ${\cal O}_5$ a
conceivable possibility is that in which the neutrino mass matrix
involves several independent sets of lepton number breaking
parameters, a situation we refer to generically as ``hybrid neutrino
masses'', as would be the case e.g. in a scheme involving interplay
between type-I and type-II seesaw \cite{Schechter:1980gr}.
Already with two contributions sourcing the neutrino mass matrix
several scenarios for neutrino mixing arising from interplay between
them can be envisaged.

In this letter we start by using an expansion of the neutrino mass
matrix in terms of the eigenvectors of the lepton mixing matrix as an
alternative model-independent parametrization of the experimentally
known values.  We show the usefulness of this treatment by studying
the constraints on the different mass matrix elements imposed by deviating from TBM,
which become evident when using this
approach due to the {\it TBM + deviations} structure the mass
matrix exhibits.
We proceed by harnessing this parametrization to analyze the
different possibilities that arise with hybrid neutrino masses, and
presenting a very appealing scenario where one of the contributions
exhibits a purely TBM form that would be well motivated by a flavor
symmetry, while the corresponding deviations, required by data are
naturally accounted for by the other contribution. In this way we
present a paradigm for neutrino masses that matches the qualitative
features required by neutrino data and in which the deviations from
TBM are interpreted as proof of the existence of hybrid neutrino
masses.


In the flavor basis (where charged leptons are diagonal) the light
neutrino mass matrix can be written as (henceforth we will denote
matrices and vectors in boldface)
\begin{equation}
  \label{eq:neu-mm}
  \boldsymbol{m_\nu}=
  \boldsymbol{U}^*\,\boldsymbol{\hat m_\nu}\,\boldsymbol{U}^\dagger\,,
\end{equation}
where $\boldsymbol{\hat
  m_\nu}=\mbox{diag}(m_{\nu_1},m_{\nu_2},m_{\nu_3})$ and with $\boldsymbol{U}$ the
lepton mixing matrix.

For any experimentally allowed point in parameter space, one can define the eigenvector
$\boldsymbol{v_i}$ associated to the eigenvalue $m_{\nu_i}$ and thus
\begin{equation}
  \label{eq:lepton-mix-eigenvectors}
  \boldsymbol{U}=\{\boldsymbol{v_1},\boldsymbol{v_2},\boldsymbol{v_3}\}
  =
  \begin{pmatrix}
    U_{11} & U_{12} & U_{13}\\
    U_{21} & U_{22} & U_{23}\\
    U_{31} & U_{32} & U_{33}
  \end{pmatrix}\,,
\end{equation}
where each $\boldsymbol{v_i}$ is a column and the neutrino mass matrix can be expressed as the outer
(tensor) product of the eigenvectors
\begin{equation}
  \label{eq:neutrino-mm-eigenvectors}
  \boldsymbol{m_\nu}=\sum_{i=1}^3m_{\nu_i}\,\boldsymbol{v_i}\otimes 
  \boldsymbol{v_i}\,.
\end{equation}
Consistency with data requires at least two eigenvectors to
be present in the above decomposition, and we refer to those cases as ``minimal''.
In the normal hierarchy a viable
minimal setup involves $\boldsymbol{v_{2,3}}$, with $\boldsymbol{v_{1,2}}$ necessary in the
inverted case.

We parametrize the mixing angles starting from TBM:
\begin{align}
  \label{eq:neutrino-mix-angles-TBM-pert}
  \sin\theta_{12}&=\sin\theta_{12}^\text{TBM}-\epsilon_{12}
  =\frac{1}{\sqrt{3}}-\epsilon_{12}\,,\\
  \label{eq:neutrino-mix-angles-TBM-pert1}
  \sin\theta_{23}&=\sin\theta_{23}^\text{TBM}-\epsilon_{23}
  =\frac{1}{\sqrt{2}}-\epsilon_{23}\,,\\
  \label{eq:neutrino-mix-angles-TBM-pert2}
  \sin\theta_{13}&=\sin\theta_{13}^\text{TBM}+\epsilon_{13}
  =\epsilon_{13}\,.
\end{align}
This is useful as according to neutrino data
\cite{Tortola:2012te,GonzalezGarcia:2012sz,Fogli:2012ua}, the
$\epsilon_{ij}$ parameters are small: at the $3\sigma$ level, for
the normal hierarchy data according to \cite{Tortola:2012te}, we extract their ranges as
\begin{align}
  \label{eq:epsilon-ranges}
  \epsilon_{12}&\subset[-0.0309,0.0577]\,,\quad \\
  \epsilon_{23}&\subset[-0.117,0.107]\,,\quad \\
  \epsilon_{13}&\subset[0.130,0.181]\,.
\end{align}
Using
this parametrization
the eigenvectors $\boldsymbol{v_i}$ can be expressed in terms of the $\epsilon_{ij}$
parameters. We write
\begin{align}
  \label{eq:eigenvectors}
  \boldsymbol{v_i}=\boldsymbol{v_i^\text{TBM}} + \boldsymbol{\varepsilon_i}\,,
\end{align}
with the TBM eigenvectors in $\boldsymbol{U_\text{TBM}}$ given by
\begin{equation}
  \label{eq:TBM-eigenvectors}
\{
  \boldsymbol{v_1}^\text{TBM},
  \boldsymbol{v_2}^\text{TBM},
  \boldsymbol{v_3}^\text{TBM} \}=
  \begin{pmatrix}
    \sqrt{2/3}  & 1/\sqrt{3} & 0\\
    -1/\sqrt{6} & 1/\sqrt{3} & -1/\sqrt{2}\\
    -1/\sqrt{6} & 1/\sqrt{3} & 1/\sqrt{2}
  \end{pmatrix}\,.
\end{equation}
The perturbation vectors $\boldsymbol{\varepsilon_i}$ can be
simplified by expanding the trigonometric functions entering in $\boldsymbol{U}$
up to second order in $\epsilon_{ij}$
\footnote{When compared with the exact expressions this approximation
  deviates at the permille level, and unitarity of $U$ is guaranteed up to corrections of order $\epsilon^3$.}.
By fixing $\delta=0$ for illustration (this does not affect the main
conclusions), they read
\begin{align}
  \label{eq:varepsilon-expressions}
  \boldsymbol{\varepsilon_1}&=
  \begin{pmatrix}
    \epsilon_{12}/\sqrt{2}\\
    \epsilon_{12}/\sqrt{2} - (\epsilon_{13}+\epsilon_{23})/\sqrt{3}\\
    \epsilon_{12}/\sqrt{2} + (\epsilon_{13}+\epsilon_{23})/\sqrt{3}
  \end{pmatrix},
  \\
  \label{eq:varepsilon-expressions1}
  \boldsymbol{\varepsilon_2}&=
  \begin{pmatrix}
    -\epsilon_{12}\\
    \epsilon_{12}/2 - \epsilon_{13}/\sqrt{6} + \sqrt{2}\epsilon_{23}/\sqrt{3}\\
    \epsilon_{12}/2 + \epsilon_{13}/\sqrt{6} - \sqrt{2}\epsilon_{23}/\sqrt{3}
  \end{pmatrix},
  \\
  \label{eq:varepsilon-expressions2}
  \boldsymbol{\varepsilon_3}&=
  \begin{pmatrix}
    -\epsilon_{13}\\
    \epsilon_{23}\\
    \epsilon_{23}
  \end{pmatrix}.
\end{align}
With the eigenvectors written as perturbations of the
$\boldsymbol{v_i}^\text{TBM}$, the neutrino mass matrix can be conveniently
interpreted as originating from a TBM structure with modifications that
are fixed whenever a given point in the corresponding experimental
data range is selected, that is to say
\begin{equation}
  \label{eq:neutrino-mm-full-eigvecs}
  \boldsymbol{m_\nu}=\sum_{i=1}^3 
  m_{\nu_i}
  \left[
    \left(
      \boldsymbol{v_i}^\text{TBM}\otimes\boldsymbol{v_i}^\text{TBM}
    \right)
    +
    \boldsymbol{{\cal V}_i}
  \right]\,,
\end{equation}
with
\begin{equation}
  \label{eq:mass-matrix-pert}
    \boldsymbol{{\cal V}_i}=
    \left[
      \left(
        \boldsymbol{v_i}^\text{TBM}\otimes \boldsymbol{\varepsilon_i}
      \right)
      +
      \left(
        \boldsymbol{\varepsilon_i}\otimes\boldsymbol{v_i}^\text{TBM}
      \right)
      +
      \left(
        \boldsymbol{\varepsilon_i}\otimes\boldsymbol{\varepsilon_i}
      \right)
    \right]\,.
\end{equation}
\begin{figure*}[t]
  \centering
  \includegraphics[width=7.5cm,height=6.1cm]{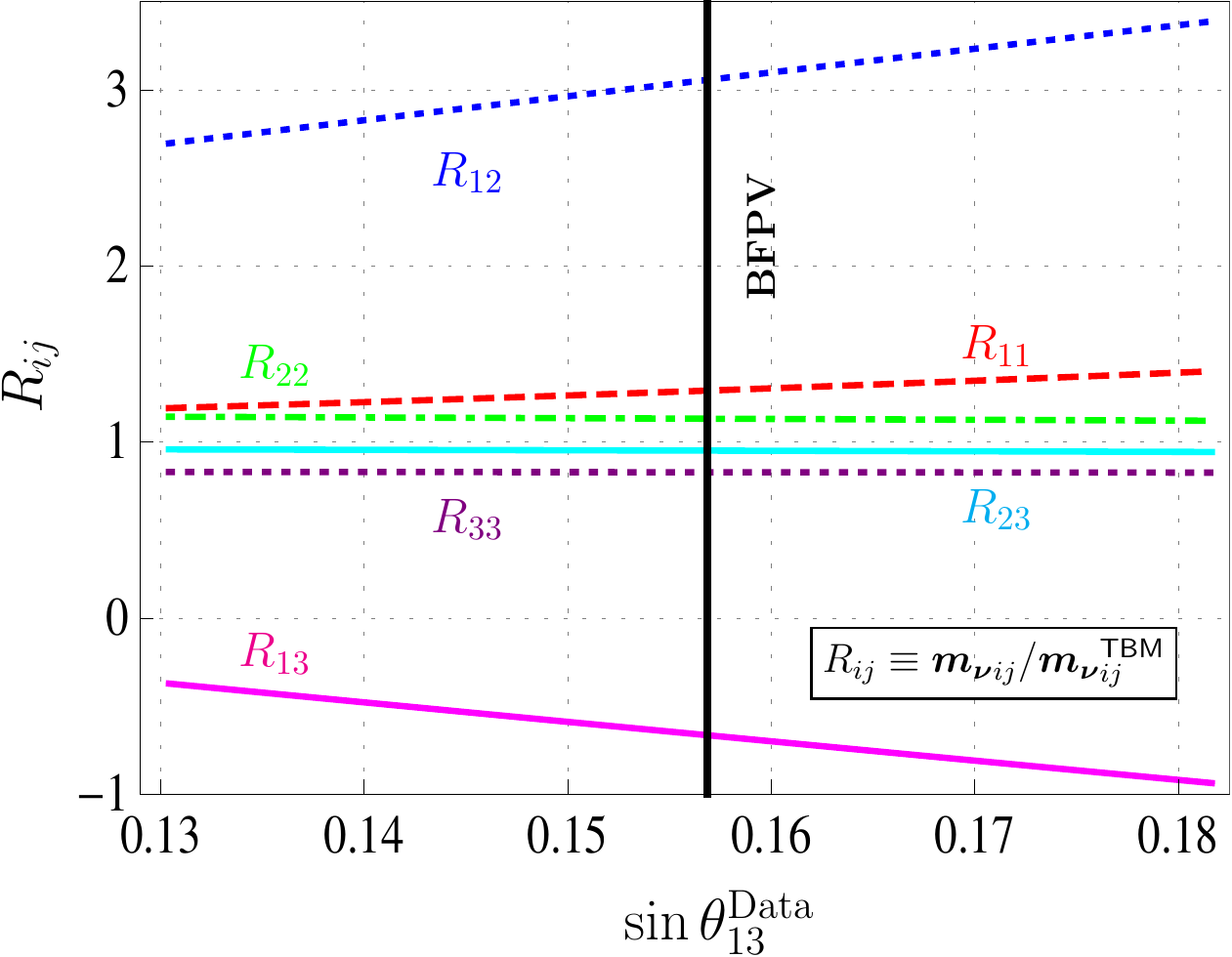}
  \hspace{0.8cm}
  \includegraphics[width=7.5cm,height=6.1cm]{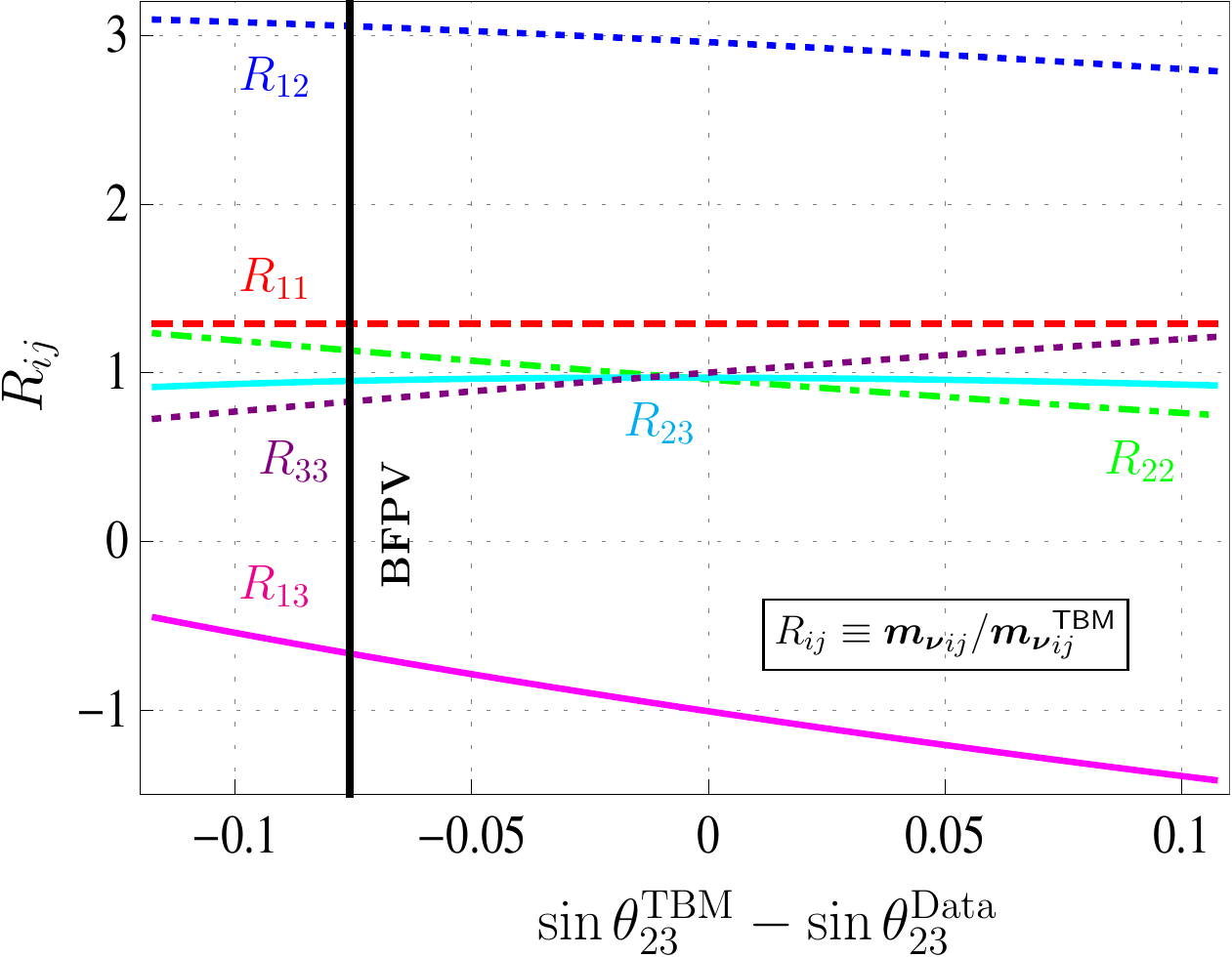}
  \caption{\it Sensitivity of the neutrino mass matrix entries,
    normalized to their TBM values, to departures from the TBM mixing
    pattern in the case of normal hierarchical light neutrino mass
    spectrum and for $\delta=0$.  Left-plot shows the dependence with
    $\sin\theta_{13}$ while the right-panel one with $\sin\theta_{23}$
    (both angles taken in their $3\sigma$ experimental range).  See
    the text for more details.}
  \label{fig:deviations-from-TBM}
\end{figure*}
It is clear that consistency with data at a certain confidence level requires some
entries of the mass matrix to significantly deviate from the TBM structure. In order to
quantify these deviations we calculated the different mass matrix
elements through eqs. (\ref{eq:neutrino-mm-full-eigvecs}) and
(\ref{eq:mass-matrix-pert}).
In fig. \ref{fig:deviations-from-TBM} we display numerical results of the different mass matrix entries (normalized to the corresponding TBM entries), $R_{ij}=m_{\nu_{ij}}/m_{\nu_{ij}}^\text{TBM}$, varying with $\theta_{13}$
and $\theta_{23}$ (the variation
with $\theta_{12}$ is weaker, so we do not display
it). We used the $3\sigma$ ranges of the angles for the
normal hierarchical spectrum according to reference \cite{Tortola:2012te}, with the remaining parameters fixed to
their BFPVs and the lightest neutrino mass set to
$10^{-3}$ eV (the results are quite insensitive to this
parameter).
These results indicate clearly that deviating from $\theta_{13}^\text{TBM}=0$
requires $m_{\nu_{12}}$ and $m_{\nu_{13}}$ to have
sizable departures from their TBM values together with small
departures in the $m_{\nu_{11}}$ entry. It can be seen in
fig. \ref{fig:deviations-from-TBM} that the other elements remain
flat, not playing a relevant role.
With $\theta_{23}$ the situation is different:
deviating from $\theta_{23}^\text{TBM}=\pi/4$ demands large deviations
from the TBM structure in $m_{\nu_{13}}$ and $m_{\nu_{12}}$, but $m_{\nu_{22}}$ and $m_{\nu_{23}}$ need also
to differ from their TBM values.  For $\theta_{12}$ the results look
similar in the sense that deviations from
$\theta_{12}^\text{TBM}$
are mostly determined by variations of $m_{\nu_{12}}$ and $m_{\nu_{13}}$ entries. Overall
in the $\delta=0$ case deviations of the neutrino mixing angles from
their TBM values require neutrino mass matrices with sizable
deviations from the TBM structure mainly in the $m_{\nu_{12}}$ and
$m_{\nu_{13}}$ elements, and this conclusion holds
independently of the neutrino mass spectrum.
We emphasize that these results apply in the flavor basis, and that we obtained these conclusions following from the useful eigenvector decomposition approach in a model-independent way.


We now apply our formalism to hybrid neutrino masses (i.e. receiving contributions
from physically distinct sources, such as different seesaw mechanisms).
In the case with two sources with superindices $A, B$
the effective light neutrino
mass matrix reads \footnote{It is straightforward to generalize to hybrid cases with more contributions.}
\begin{equation}
  \label{eq:nmm-several-sources}
  \boldsymbol{m_\nu}=\boldsymbol{m_\nu^{(A)}} 
  + \boldsymbol{m_\nu^{(B)}}\,.
\end{equation}
From equation (\ref{eq:neu-mm}) we have
\begin{equation}
  \label{eq:diagonalizing-ApB}
  \boldsymbol{\hat m_\nu}=
  \boldsymbol{U}^T
  \left(
    \boldsymbol{m_\nu^{(A)}}
    +\boldsymbol{m_\nu^{(B)}}
  \right)
  \boldsymbol{U}\,.
\end{equation}
and in general situation $\boldsymbol{U}$ diagonalizes the sum but not
the individual matrices $\boldsymbol{m_\nu^{(A,B)}}$. The
contributions must add up to (\ref{eq:neutrino-mm-full-eigvecs}) but
their individual structure needs not be determined by the eigenvectors
of $\boldsymbol{U}$.
The most general decomposition can be written as
\begin{equation}
  \label{eq:general-mA-mB-decomposition}
  \boldsymbol{m_\nu^{(X)}}=\sum_i 
  \left[
    m_{\nu_i}^{(X)}
    \boldsymbol{v_i}^\text{TBM}\otimes \boldsymbol{v_i}^\text{TBM}
    +
    \delta m_{\nu_i}^{(X)}\boldsymbol{{\cal V}_i}
  \right]\,,
\end{equation}
where $X=A,B$ and by definition, (\ref{eq:diagonalizing-ApB}) requires $m_{\nu_i}^{(A)} +
m_{\nu_i}^{(B)}=\delta m_{\nu_i}^{(A)} + \delta
m_{\nu_i}^{(B)} =m_{\nu_i}$.
We stress that any realization of hybrid neutrino masses can be defined
according to the terms entering in each contribution.
Consistency requires that when combining $\boldsymbol{m_\nu^{(A,B)}}$ through (\ref{eq:nmm-several-sources})
the eigenvectors entering in the full mass matrix sum up
to~(\ref{eq:eigenvectors}), or in other words that the matrix
associated with the generation index $i$ has at the end a structure
like (\ref{eq:neutrino-mm-full-eigvecs}).
Regardless of which eigenvectors appear in each individual contribution, the orthogonality
relation $\boldsymbol{v_i}\cdot \boldsymbol{v_j}=\delta_{ij}$ guarantees the lepton
mixing matrix diagonalizes the resulting $\boldsymbol{m_\nu}$ (due to $\boldsymbol{v_i}$ being approximate, this holds up to
corrections at most of order
$v_{kl}^\text{TBM}\epsilon_{ij}\sim 10^{-1}$).
Indeed, it is useful to apply the eigenvector decomposition to each contribution as it is made clear that the only way for the eigenvectors building either $\boldsymbol{m_\nu^{(X)}}$ to appear in $U$ unchanged is if they are already orthogonal, which in general will not be the case.
For illustration, we consider a minimal setup (minimal in terms of the number of parameters defining the full neutrino mass matrix)
\begin{align}
  \label{eq:minimal-setup}
  \boldsymbol{m_\nu^{(A)}}&=m_{\nu_2}\boldsymbol{v_2}^\text{TBM}\otimes 
  \boldsymbol{v_2}^\text{TBM} \,,\\
  \boldsymbol{m_\nu^{(B)}}&=m_{\nu_3}\boldsymbol{v_3}\otimes 
  \boldsymbol{v_3} + m_{\nu_2}\boldsymbol{{\cal V}_2}\,.
\end{align}
In this case the vector $\boldsymbol{v_2}$ becomes ``completed''
through the combination of $\boldsymbol{m_\nu^{(A)}}$ and the second
term in $\boldsymbol{m_\nu^{(B)}}$. This can be seen explicitly as it results in a particular case of (\ref{eq:neutrino-mm-full-eigvecs}).
Minimal setups are appealing as there are only 6 defining parameters (with a phase in addition to the 3 $\epsilon_{ij}$), so there is no room for arbitrariness on the parameter space in order to match the neutrino oscillation observables ($\Delta m_{21,32}$, $\theta_{ij}$
and $\delta$).
Although at the expense of introducing more parameters, going beyond minimal cases can lead to other possibilities.
A classification according to the number of eigenvectors included in each mechanism can be done in analogy to the one shown in ref \cite{AristizabalSierra:2011ab} for the exact TBM pattern.

Another appealing setup (minimal or not) is one where one of the structures (e.g. $\boldsymbol{m_\nu^{(A)}}$) is chosen to
involve only TBM eigenvectors (as in
(\ref{eq:minimal-setup})). In non-minimal scenarios of this type we can have the TBM pattern arise solely from one
of the contributions while the other entirely accounts for the observed deviations. This corresponds to setting $\delta m_{\nu_i}^{(A)}=m_{\nu_i}^{(B)}=0$ in (\ref{eq:general-mA-mB-decomposition}):
\begin{align}
  \label{eq:TBM+deviations-case}
  \boldsymbol{m_\nu^{(A)}}&=\sum_i m_{\nu_i}
  \boldsymbol{v_i}^\text{TBM}\otimes\boldsymbol{v_i}^\text{TBM}\,, \nonumber\\
  \boldsymbol{m_\nu^{(B)}}&=\sum_i m_{\nu_i} \boldsymbol{{\cal V}_i}\,,  
\end{align}
which is very suggestive that the experimentally observed small deviations from
the TBM pattern may be interpreted as a hint that
nature is described by hybrid neutrino masses. With the measured deviations from TBM such a scenario is extremely natural:
in general we expect the eigenvectors associated with different mechanisms to not be orthogonal, and the smallness of the deviations would simply be due to a moderate hierarchy in the scales associated with each mechanism.

In conclusion, we parametrized the neutrino mixing angles by
small perturbations of the TBM pattern,
and expressed the eigenvectors of the neutrino mass matrix (in the flavor basis) in terms
of a dominant TBM structure with
small perturbations.
To very good approximation the TBM deviations are simple eigenvectors depending linearly in the deviations of the mixing angles.
This approach was used first to clarify which neutrino mass matrix elements are associated with the deviations of each angle.
We then applied the same approach to ``hybrid'' neutrino mass matrices, where it conveniently describes how the eigenvectors from different sources
of neutrino masses combine into the observed mixing.
We identified some particularly appealing cases, starting with the minimal ones where a small number of parameters makes the scheme predictive and then considering cases where one of the sources had a mass matrix with the exact TBM form. We argued that given the data, the latter are extremely natural, 
and claim that neutrino mixing data might therefore be the first hint of the presence of several mechanisms generating neutrino masses in nature. 
Such a framework is a novel perspective for neutrino mixing which deserves
further theoretical
and phenomenological scrutiny.

Acknowledgments: The work of I.d.M.V. was supported by DFG grant PA
803/6-1 and partially through PTDC/FIS/098188/2008. DAS is supported
by the Belgian FNRS agency through a ``Charg\'e de recherche''
contract.

\end{document}